\documentclass[a4, 11pt]{article}
\usepackage[T1]{fontenc}

\usepackage{amssymb}        
\usepackage{latexsym}
\usepackage{times}
\usepackage{array}

\usepackage{graphicx}

\author{}

\date{}

\title{Maximum ATSP with Weights Zero and One via Half-Edges}

\author{Katarzyna Paluch\\
Institute of Computer Science,  University of Wroc{\l}aw \\
{\tt abraka@cs.uni.wroc.pl}}

\newcommand{\dowod}{\noindent{\bf Proof.~}}
\newcommand{\koniec}{\hfill $\Box$\\[.1ex]}
\newtheorem{fact}{Fact}

\newtheorem{lemma}{Lemma}
\newtheorem{theorem}{Theorem}
\newtheorem{corollary}{Corollary}
\newtheorem{definition}{Definition}
\newtheorem{observation}{Observation}

\begin{document}

\maketitle
\thispagestyle{empty}
\begin{abstract}
We present a fast combinatorial $3/4$-approximation algorithm for the maximum asymmetric TSP with weights zero and one. The approximation factor of this algorithm matches the currently best one given by Bl{\"a}ser  in 2004 and  based on linear programming.
Our algorithm first computes a maximum size matching and a maximum weight cycle cover without certain cycles  of length two but possibly with {\em half-edges} - a half-edge of a given edge $e$ is informally speaking a half of $e$ that contains one of the endpoints of $e$.
Then from the computed matching and cycle cover it extracts a set of paths, whose weight is large enough to be able to construct a traveling salesman tour with the claimed guarantee.


\end{abstract}

\section{Introduction}
We study the maximum asymmetric traveling salesman problem with weights zero and one (Max (0,1)-ATSP), which is defined as follows. Given a complete loopless directed graph $G$ with edge weights zero and one, we wish to compute a traveling salesman tour of  maximum weight.
Traveling salesman problems with weights one and two are an important special case of traveling salesman problems with triangle inequality.  Max (0,1)-ATSP is connected to Min (1,2)-ATSP (the minimum asymmetric traveling salesman problem wth weights one and two) in the following way. 
It has been shown by Vishvanathan \cite{Vish} that a $(1-\alpha)$-approximation algorithm for Max (0,1)-ATSP yields a $(1+\alpha)$-approximation algorithm for Min (1,2)-ATSP by replacing weight two with weight zero. 

Approximating Max (0,1)-ATSP with the ratio $1/2$ is easy -- it suffices to compute a maximum weight matching of the graph $G$  and patch the edges arbitrarily into a tour. The first nontrivial approximation of Max (0,1)-ATSP was given by Vishvanathan \cite{Vish} and has the approximation factor $7/12$. It was improved on by Kosaraju, Park, and Stein \cite{KPS}  in 1994, who  gave a $48/63$-approximation algorithm that also worked for Max ATSP with arbitrary nonnegative weights. Later, Bl{\"a}ser and Siebert \cite{BS} obtained a $4/3$-approximation algorithm for Min (1,2)-ATSP , which can be modified to give a $2/3$-approximation algorithm for Max (0,1)-ATSP.  $2/3$-approximation algorithms are also known for the general Max ATSP and have been given in \cite{KLSS} and \cite{PEZ}.
The currently best published approximation algorithm for Max (0,1)-ATSP  achieving ratio $3/4$ is due to   Bl{\"a}ser  \cite{B34}. It uses linear programming to obtain a multigraph $G_M$ of weight at least $3/2$ times the weight of an optimal traveling salesman tour (OPT) such that $G_M$ can be {\em path-2-colored}. A multigraph is called {\em path-2-colorable} if its edges can be colored with two colors so that each color class consists of vertex-disjoint paths. The   algorithm by  Bl{\"a}ser has a polynomial running time but the degree of the polynomial is high.  A $3/4$-approximation algorithm for  Max ATSP with arbitrary nonnegative weights has been given in \cite{Pal34}. The presented here algorithm for Max (0,1)-ATSP is much simpler than the one in \cite{Pal34}.

Karpinski and Schmied have shown in \cite{Karp} that it is NP-hard to approximate Min (1,2)-ATSP with an approximation factor less than $207/206$ and for the general Max ATSP  that it is NP-hard to obtain an approximation better than $203/204$.

{\bf Our approach and results} We present a simple combinatorial $3/4$-approximation algorithm for Max (0,1)-ATSP.  First we compute a maximum weight matching $M_{max}$ of $G$. By a matching of $G$ we mean any vertex-disjoint collection of edges.  The weight of $M_{max}$ is clearly at least OPT/2, where OPT denotes the weight on an optimal tour.
Next, we compute a maximum weight {\em cycle cover} that {\em evades the matching $M_{max}$}. A {\em cycle cover} of a directed graph is such a collection of directed cycles that each vertex belongs to exactly one cycle of the collection.  A {\em cycle cover  of a graph $G$ that evades a matching $M$ } is a cycle cover of $G$ which does not contain any length two cycle (called a {\em  2-cycle}) going through two vertices that are connected by some edge of $M$ but it may contain {\em half-edges} - a half-edge of a given edge $e$ is informally speaking a half of $e$ that contains one of the endpoints of $e$. Half-edges have already been introduced in \cite{PEZ}.The task of finding a maximum weight cycle cover  $C_{max}$ that evades a matching $M$ 
can be reduced to finding a maximum size matching in an appropriately constructed graph.  The weight of $C_{max}$ is an upper bound on OPT. Further on we show that a maximum weight matching $M_{max}$ and a maximum weight cycle cover that evades $M_{max}$ can be easily transformed into a path-2-colorable multigraph. For completeness we give also our own linear time procedure of path-2-coloring. This method takes advantage of the fact that the edge weights are zero and one.  A more general algorithm for path-2-coloring that runs in $O(n^3)$ has been given in \cite{B34}.

This way the main results of this paper can be stated as

\begin{theorem}

There exists a combinatorial $3/4$-approximation algorithm for Max (0,1)-ATSP. Its running time is $O(n^{1/2}m)$, where $n$ and $m$ denote the number of respectively vertices and edges of weight one  in the graph.
\end{theorem}

\begin{corollary}
There exists a combinatorial $5/4$-approximation algorithm for Min (1,2)-ATSP. Its running time is $O(n^{1/2}m)$
\end{corollary}

\section{Cycle cover that evades  matching $M$}

The algorithm for Max (0,1)-ATSP starts from computing a maximum weight perfect matching $M_{max}$ of $G$. By a {\bf \em 0-edge} and a {\bf \em 1-edge} we will mean an edge of weight, respectively, zero or one.  By $G_1$ we  denote the subgraph of $G$ consisting of all 1-edges of $G$. In order to obtain a maximum weight perfect matching $M_{max}$ of $G$, it is enough
to compute a maximum size matching $M_1$  in $G_1$ and, if necessary, complete it arbitrarily with 0-edges so that the resulting matching is perfect.

Next, we would like to find a maximum weight cycle cover of $G$ that does not contain any $2$-cycle in $G_1$, whose one edge  belongs to $M_{max}$.
Since computing such  a cycle cover is NP-hard, which follows from a similar result proved by Bl{\"a}ser \cite{}, we are going to relax the notion of a cycle cover and allow it to contain {\bf \em half-edges}  - a half-edge of edge $(u,v)$ is informally speaking
``half of the edge $(u,v)$ that contains either a head or a tail of $(u,v)$''.

Now, we are going to give a precise definition of a cycle cover that evades a matching $M$.   We say that a $2$-cycle $c$  in $G_1$ is {\bf \em $M$-hit} if one of the edges of $c$ belongs to $M$.  We introduce a graph $\tilde G$.  $\tilde G =(\tilde V, \tilde E)$  is the graph obtained from $G$ by splitting 
 each edge  $(u,v) $  belonging to a $M$-hit $2$-cycle of $G_1$ with a vertex $x_{(u,v)}$  into two edges $(u,x_{(u,v)})$ and $(x_{(u,v)},v)$, each with weight $\frac 12 w(u,v)$, where  $w(u,v)$ denotes the weight of the edge $(u,v)$.  Each of the edges $(u, x_{(u,v)}),  (x_{(u,v)},v)$ is called
{\bf \em a half-edge (of $(u,v)$)}.   For any subset of edges $E' \subseteq E$ by $w(E')$ we mean $\sum_{e \in E'} w(e)$.

\begin{definition}\label{rel2}
A {\bf \em  cycle cover   that evades a matching $M$}  is a subset $\tilde C\subseteq \tilde E$ such that
\begin{itemize}
\item[(i)]
each vertex in $V$ has exactly one outgoing and one incoming edge in $\tilde C$;

\item[(ii)]
for each $M$-hit $2$-cycle of $G_1$ connecting vertices $u$ and $v$ $\tilde C$ contains either zero or  two edges from 
\newline $\{(u, x_{(u,v)}), (x_{(u,v)}, v), (v, x_{(v,u)}), (x_{(v,u)}, u)\}$.  Moreover, if $\tilde C$ contains  only one half-edge of $(u,v)$ , then it also contains one half-edge of $(v,u)$, and  one of these half-edges  is incident with $u$ and the other with $v$.
\end{itemize}
\end{definition}

To compute a  cycle cover  $C_1$  that evades  $M_{max}$  we construct the following undirected  graph $G'=(V',E')$.
For each vertex $v$ of $G$ we add two vertices $v_{in}, v_{out}$ to $V'$. For each edge $(u,v) \in E$
we add vertices $e^1_{uv}, e^2_{uv}$, an edge  $(e^1_{uv}, e^2_{uv})$ of weight $0$ and edges $(u_{out}, e^1_{uv}), (v_{in}, e^2_{uv})$, each of weight $\frac{1}{2}w(u,v)$.
Next we build so-called gadgets.

For each  $M$-hit $2$-cycle in $G_1$ on vertices $u$ and $v$  we add vertices $a_{\{u,v\}}, b_{\{u,v\}}$ 
and edges $(a_{\{u,v\}}, e^1_{uv}), \\ (a_{\{u,v\}}, e^2_{vu}, (b_{\{u,v\}}, e^1_{vu}), (b_{\{u,v\}}, e^2_{uv})$ having weight $0$. 


\begin{theorem}
Any perfect matching of $G'$ yields  a  cycle cover  $C_1$ that evades $M_{max}$.
A maximum weight perfect matching of $G'$ yields a  cycle cover $C_{max}$ that evades $M_{max}$  such that $w(C_{max}) \geq OPT$.
\end{theorem}
\dowod
The proof of the first statement is very similar to the proof of Lemma 2 in \cite{PEZ}.
The second statement follows from the fact that  a traveling salesman tour is also a cycle cover that evades $M_{max}$.
\koniec

A cycle cover that evades a matching $M$ consists of directed cycles and/or directed paths, where each of the directed paths begins and ends with a half-edge. In the following by  a half-edge of a cycle cover $C$  we  will mean such a half-edge  of a certain edge $e$  contained in  $C$ that $C$ contains only  one half-edge of $e$.
From a matching $M_{max}$ and a maximum weight cycle cover $C_{max}$ that evades $M_{max}$ we build a multigraph $G_m$ as follows. Basically $G_m$ consists of one copy of $M_{max}$ and one copy of $C_{max}$. However, we do not want $G_m$ to contain half-edges. Therefore we modify  $C_{max}$ by replacing  each pair of half-edges of edges connecting vertices $u$ and $v$ that are contained in $C_{max}$  with an edge $(u,v)$, if $M_{max}$ contains $(v,u)$  and  otherwise with an  edge $(v,u)$. As a result $G_m$ contains a $2$-cycle on each such pair of vertices $u,v$.  After this modification $C_{max}$ contains only whole edges and may contain directed paths with a common endpoint i.e.,  some vertices may have indegree two and outdegree zero  or vice versa.  However, the overall weight  of $C_{max}$ is unchanged. Now, $G_m$ is going to contain two copies of an edge $e$ if $e$ belongs both to $M_{max}$ and $C_{max}$ and one copy of an edge $e$ if $e$ belongs either to $M_{max}$ or to $C_{max}$.
This way we obtain a multigraph that satisfies the following conditions:
\begin{itemize}
\item each vertex in $G_m$ has degree three,
\item each vertex in $G_m$ has indegree at most two and outdegree at most two,
\item for each pair of vertices $u$ and $v$, $G_m$ contains at most two edges connecting $u$ and $v$.
\end{itemize}

In \cite{B34} Bl{\"a}ser  shows how to slightly modify such a multigraph so that it has the same number of 1-edges and is path-2-colrable. Path-2-coloring of the modified graph is based on a variant of the path-$2$-coloring lemma given by Lewenstein and Sviridenko \cite{LS},
which in turn is a reduction to the path-$2$-coloring lemma of Kosaraju, Park, and Stein, whose proof was given in \cite{B1}. The running time of the path-$2$-coloring algorithm  is 
 $O(n^3)$.

If the number of vertices in the graph is odd, then the above approach does not give a $3/4$-approximation. We can either add a new additional vertex, that is connected to every other vertex by a 0-edge and obtain a $3/4(1-1/n)$-approximation, or
guess two consecutive edges of an optimal traveling salesman tour and contract them. In the latter case, the running time of the algorithm becomes $O(n^{5/2}m)$.

\section{Path-2-coloring}

From $G_m$ we are going to obtain another multigraph that contains the same number of 1-edges as $G_m$ and and additionally allows a simple method of path-2-coloring.

First we deal with $2$-cycles on cycles and paths of $C_{max}$. For any 1-edge $e=(u,v)$ contained in a cycle $c$ of $C_{max}$  such that $M_{max}$ contains a $1$-edge $e'=(v,u)$, we replace the edge $e'$ with another copy of $e$.
Similarly,  for any 1-edge $e=(u,v)$ contained in a path $p$ of $C_{max}$  such that $e$ is not an ending edge of $p$ and $M_{max}$ contains a $1$-edge $e'=(v,u)$, we replace the edge $e'$ with another copy of $e$.
So far, clearly, we have not diminished the number of 1-edges contained in $G_m$. Next, we are going to discard all $0$-edges from $G_m$.  This way, some cycles of $C_{max}$ may disintegrate into paths and some paths of $C_{max}$ may also give rise to shorter
or new paths. In what follows, by a cycle of $C_{max}$ we will mean a cycle of $C_{max}$ consisting solely of 1-edges and by a path of $C_{max}$  we will mean a maximal (under inclusion) directed path, whose every edge  belongs to $C_{max}$ and has weight one.

Let $e=(u,v)$ be an edge,  $c$ a cycle  and $p$ a path of $C_{max}$.  Then we say that $e$ is an {\bf \em inray of $c$ (corr. $p$)} if $u \notin c$ and $v \in c$ (corr.   $u \notin p$ and $v \in p$). If  $u \in c$ and $v \notin c$ (corr.  $u \in p$ and $v \notin p$), then we say that $e$ is an {\bf \em outray of $c$ (corr. $p$)}.  A {\bf \em ray of $c$ ($p$)} is any inray or outray of $c$ ($p$).   If both  endpoints of $e$ belong to $c$ (corr. $p$) and $e$ does not belong to $c$ (corr. $p$), then $e$ is called a {\bf \em chord} of $c$ (corr. $p$). If $e$ is a copy of some edge belonging to $c$ (corr. $p$), then $e$ is called
an {\bf \em ichord}.

  Let us notice that any $2$-cycle which is present at this stage of $G_m$ is either a $2$-cycle of $C_{max}$ or a $2$-cycle obtained from a pair of half-edges of $C_{max}$ and an edge of $M_{max}$. Now,  
if $c$ is a $2$-cycle of $C_{max}$ on vertices $u$ and $v$  that has an inray incident to $u$ and an outray incident to $v$, then we replace the edge $(v,u)$ with another copy of edge $(u,v)$  and shrink the two copies of an edge $(u,v)$ into a single vertex. Every remaining $2$-cycle of $C_{max}$  or a $2$-cycle obtained from a pair of half-edges of $C_{max}$ and an edge of $M_{max}$ is also shrunk into a single vertex. 

Let us call the multigraph obtained from $G_m$ by shrinking all such  $2$-cycles $G'_m$. We make the following observation.
\begin{observation}
From any path-2-coloring of $G'_m$ we can obtain a path-2-coloring of $G_m$ without changing the color of any edge of $G'_m$.
\end{observation}

Next we are going to further  flip some of the edges of $G'_m$ to make the task of its path-2-coloring  very easy.

For each cycle $c$ of $C_{max}$ we are going to flip either   its inrays and chords or outrays and chords so that $c$ has either  only outrays and ichords  or only inrays and ichords.
Let $c$ be any cycle of $C_{max}$. Let us notice that its length is at least three.  Suppose that the number of inrays of $c$ is not smaller than the number of outrays of $c$. Then the inrays are left as they are and the outrays and chords of $c$ are flipped so that they become  ichords,  i.e. each inray and chord of $c$ is replaced 
 with a copy of some edge of $c$. The flipping is done in such a way that the indegree and outdegree of each vertex of $c$ is at most two. Now, however,  it may happen that some vertex of $c$ has both indegree and outdegree equal to two. More precisely the process of flipping looks as follows. Let $E_c$ be a subset of edges of $c$ such that an edge  $(u,v)$ of $c$  belongs to  $E_c$ if no inray of $c$ is incident with $v$. The number of edges in $E_c$ is not smaller than the number of outrays and chords of $c$. Moreover, the number of outrays and chords of $c$ is not bigger than $|c|-2$, where $|c|$ denotes the length of $c$. It is so because the number of chords of $c$ is not greater than $|c|/2$  and the number of outrays of $c$ is not bigger  than the number of inrays of $c$.  Each chord and outray of $c$ is replaced with a copy of some edge of $E_c$.
If the number of outrays of $c$ outnumbers the number of inrays of $c$, then we flip the inrays and chords of $c$ so that they become ichords in an analogous way as above.

\begin{fact}\label{cykl}
Let $c$ be any cycle that has either  only inrays and/or ichords    or only outrays and/or ichords. Moreover, (1) the number of rays of $c$ is at least two or $c$ has at most $|c|-2$  ichords and (2) the indegree and outdegree of each vertex of $c$ is at most two.
If $c$  has at least two rays, then  it is possible to path-$2$-color the edges and ichords of $c$ if  two rays of $c$ are colored differently. If $c$ has at most one ray, then it is always possible to path-$2$-color the edges and ichords of $c$.
\end{fact}
\dowod
Any two copies of the same edge must be colored differently. Similarly any two outgoing edges of some vertex of $c$ or any two incoming edges of some vertex of $c$ must be colored differently. If $c$ has two  rays that are colored differently, then it follows that
we are unable to create a monochromatic cycle out of the edges or ichords of $c$.  If $c$ has exactly one ray colored with, say $1$, then we must see to it that not for every edge $(u,v)$ of $c$ it is that at least one copy of $(u,v)$  is colored with $2$.
Since $c$ has at most $|c|-2$ ichords, there exists an edge $e$ of $c$ such that $G'_m$ contains only one copy of $e$ and which can be colored with $1$. If $c$ has no rays, then we can easily path-$2$-color its edges and ichords.
\koniec

The situation with paths is a little bit more complicated. We are going to distinguish paths that are {\bf \em bound} and {\bf \em free}. A path of $C_{max}$ is said to be bound if it shares at least one of its endpoints with another path of $C_{max}$. A path of $C_{max}$
that is not bound is said to be free. A bound path can be {\bf \em 1-bound} -- if exactly one of its endpoints is also an endpoint of another path of $C_{max}$  or {\bf \em 2-bound} -- if each of its endpoints is an endpoint of another path of $C_{max}$.
We say that an edge $e=(u,v)$ of $p$ of $C_{max}$  is a {\bf \em rayter}  if $u$ is incident with an outray of  $p$ and $v$ is incident with an inray of $p$.

We are going to flip the rays and chords of each bound path $p$  in such a way that besides possible  ichords  $p$ either  has  at most one ray  or exactly exactly two rays incident to a rayter. As for free paths we are going to flip the rays and chords of each free path $p$  in such a way 
that besides possible ichords $p$ either has only inrays or only outrays or exactly two rays incident to a  rayter.

Let $p$ be any path of $C_{max}$ with endpoints $u$ and $v$.  By $|p|$ we denote the length of $p$, i.e., the number of edges of $p$.  An endpoint of $p$ which is not an endpoint of any other path of $C_{max}$ is said to be a {\bf \em border vertex} of $p$.
If an endpoint $u$ of $p$ belongs also to some other path of $C_{max}$, then the  edge of $p$ incident to $u$ is called a  {\bf \em border edge} of $p$. The endpoint of a border edge of $p$ that is not an endpoint of any  path of $C_{max}$ different from $p$ is also called
a {\bf \em border vertex} of $p$. It may happen that a path $p$ of $C_{max}$ does not have any border vertex --  if $|p|=1$ and both endpoints of $p$ belong also to some other path(s) of $C_{max}$. We say that a path $p$ has a {\bf \em good ray}  if it has a ray $e$ incident to a border vertex $v$ of $p$ such that either (1) $v$ is an endpoint of $p$ and $e$ together with $p$ form a directed path of length $|p|+1$ or (2)  $v$ is not an endpoint of $p$  and $e$  forms a directed path of length two with $e'$, where $e'$  is an edge of $p$ incident to $v$  and is not a border edge of $p$. For example, let $p$ be a $2$-bound path $(u, v_1, v_2, v)$ directed from $u$ to $v$ and suppose that $p$ has a ray  $e=(v_2, v_3)$ . Then $e$ is a good ray of $p$.  
Let us notice that the maximum number of edges of $M_{max}$ incident to a  path $p$ of $C_{max}$  is: (1) $|p|-1$, if $p$ is $2$-bound, (2) $|p|$, if $p$ is $1$-bound and (3) $|p|+1$, if $p$ is free.  It is so because no edge of $M_{max}$ is incident to a vertex which is an endpoint of two different paths of $C_{max}$  -- because such an endpoint is in fact a shrunk $2$-cycle.

The flipping of rays and ichords of paths proceeds as follows.  If  the number of edges of $M_{max}$  incident to a given  path $p$  is (1)  fewer than  $|p|-1$ and $p$ is $2$-bound or (2) fewer than $|p|$ and $p$ is $1$-bound or (3) fewer than $|p|+1$ and $p$ is free,
then we flip all chords and rays of $p$ so that they become ichords and so that no ichord is a copy  of any border edge of $p$. (Also, of course, no edge of $p$ is allowed  to occur in more than two copies.)  Otherwise, if a path $p$ has a good ray, we leave any one good ray of $p$ as it is and flip all the other rays and chords of $p$  so that they become ichords and  no ichord is a copy  of any border edge of $p$. In the reamining case, we leave some  two rays of $p$ that are incident to a rayter and flip the rest of rays and chords of $p$ so that they become ichords.

Suppose that $e_1$ and $e_2$ are good rays of paths $p_1, p_2$  having a common endpoint $u$ such that both $e_1$ and $e_2$ is incident to the border edge (of respectively $p_1$ or $p_2$)  incident with $u$. Then the rays $e_1$ and $e_2$ are  said to be {\bf \em allied}.

We make the following two observations.

\begin{fact}
In any path-$2$-coloring of $G'_m$ the rays incident to the same rayter are colored with the same color.
\end{fact}
\dowod
Let $e=(u,v)$ be a rayter of $p$. Then in any path-$2$-coloring of $G'_m$ the edge $e$ must be colored with a different color than an outray of $p$ incident to $u$ and also with a different color than an inray of $p$ incident to $v$. Since there are only two colors,
it follows that the rays incident to $e$ must be colored with the same color.
\koniec

\begin{fact}
In any path-$2$-coloring of $G'_m$ the allied rays are colored with different colors.
\end{fact}
\dowod
Let $v$ be a vertex which is an endpoint of two different paths $p_1, p_2$ of $C_{max}$ and let $e_1, e_2$ be two border edges incident to $v$. Then, clearly $e_1$ and $e_2$ must be colored with different colors as either both are the incoming edges of $v$ or both are
the outgoing edges of $v$. The ray incident to $e_1$ must be colored differently than $e_1$.  Similarly the ray incident to $e_2$ must be colored differently than $e_2$.
\koniec

After all the flipping, the multigraph $G'_m$ is quite easy to path-$2$-color. In fact, it suffices to appropriately  color the rays and then the coloring of the rest of the edges is straightforward. From the rays in $G'_m$ we build the following graph $H$. At the beginning $H$ has the same vertex set as $G'_m$ and contains all the rays in $G'_m$, i.e., $(u,v)$ is an edge in $H$
if and only if $(u,v)$ is a ray of some path or cycle of $C_{max}$  in $G'_m$ after  the flipping. Next, for each cycle $c$ of $C_{max}$ we choose two arbitrary rays $e_1, e_2$  of $c$  and glue together their endpoints belonging to $c$ i.e., if $u_1 \in e_1 \cap c$ and $u_2 \in e_2 \cap c$, then we replace $u_1$ and $u_2$ with one vertex and as a result $e_1$ and $e_2$ have (at least)  one common endpoint. Further, each pair of rays incident to the same rayter is replaced with one edge as follows. Let $e_1=(u_1, v_1), e_2=(u_2, v_2)$  be a pair of rays  incident to some edge $e=(u_2,v_1)$  in $G'_m$. Then $e_1, e_2$ are replaced  in $H$ with one edge $e=(u_1, v_2)$. Such replacements are done exhaustively.  
We also glue together the endpoints of certain pairs of good rays. Suppose that $e_1$ and $e_2$ are allied  rays of paths $p_1, p_2$. Then we glue together the endpoint of $e_1$ belonging to $p_1$ with the endpoint of $e_2$ belonging to $p_2$.

At this stage, ignoring the directions $H$ consists of paths,  cycles and isolated vertices, i.e. each vertex is either isolated or belongs to exactly one path or cycle.  Moreover, if some cycle in $H$ is of odd length, then it contains at least two consecutive edges that form a directed path. We color the edges of each path and cycle of $H$ alternately with $1$ and $2$ in such a way that no two incoming edges of any vertex are colored with the same color or no two  outgoing edge of any vertex are colored with the same color. In other words, we path-$2$-color $H$.

\begin{lemma}
Any path-$2$-coloring of $H$ can be extended to a path-$2$-coloring of $G'_m$.
\end{lemma}
\dowod
Each ray in $G'_m$ is colored with the same color as  in $H$. In the  case when  some edge $e$ in $H$ was obtained from several rays in $G'_m$, each such ray in $G'_m$ is colored in the same way as $e$ in $H$. Thus, by the way we constructed $H$,  each pair of rays incient to one rayter  is colored in the same way,
allied rays are colored with different colors and for each cycle $c$ of $C_{max}$ that has at least two rays, there exist two rays of $c$ colored differently.  By Fact \ref{cykl} we already know how to color the edges and ichords of each cycle of $C_{max}$.
Any edge $e=(u,v)$  of any path of $C_{max}$ which is incident to an outray $r_1$ incident to $u$ is colored differently than $r_1$. Similarly any edge $e=(u,v)$  of any path of $C_{max}$ which is incident to an inray $r_2$ incident to $v$ is colored differently than $r_2$. 
Also two border edges of two different paths of $C_{max}$ incident to the same vertex are colored differently. Two copies of the same edge are clearly colored differently. The remaining edges can be colored arbitrarily.

\koniec

\end{document}